\documentclass[preprint,aps,pre]{revtex4}
\usepackage[dvips]{graphicx}
\usepackage{latexsym}
\usepackage{epsfig}
\usepackage{bm}
\usepackage{times,psfrag,subfigure}        
\begin{document}

\title{Controlling drop size and polydispersity using chemically
patterned surfaces}
\author{H. Kusumaatmaja}
\author{J. M. Yeomans}
\affiliation{The Rudolf Peierls Centre for Theoretical Physics, Oxford University, 1 Keble Road, Oxford OX1 3NP, U.K.}
\date{\today}

\begin{abstract}
We explore numerically the feasibility of
using chemical patterning to control the size and polydispersity of
micron-scale drops. The simulations suggest that it is possible to 
sort drops by size or wetting properties by using an array of 
hydrophilic stripes of different widths. We also demonstrate that monodisperse drops can be generated by exploiting 
the pinning of a drop on a hydrophilic stripe. Our results follow 
from using a lattice Boltzmann algorithm to solve the hydrodynamic 
equations of motion of the drops and demonstrate the applicability 
of this approach as a design tool for micofluidic devices with chemically
patterned surfaces.

\end{abstract}
\maketitle


\section{Introduction}

In recent years advances in patterning techniques, as well as in
our understanding of electrowetting, have made it
increasingly feasible to construct surfaces with regions of different
wettability on micron length scales. The behaviour of fluids spreading
and moving across such surfaces is extremely rich, and is only just
beginning to be explored \cite{Kusumaatmaja1,Leopoldes1,Dupuis2,Dupuis1,Lenz1,Lipowsky1,Lipowsky2,
Darhuber1,Darhuber2,Mugele1,Mugele2}. Biosystems have evolved to use hydrophobic
and hydrophilic patches to direct the motion of fluids at surfaces
\cite{Blossey1,Parker1}. Examples are desert beetles, whose patterned backs
help them to collect dew from the wind, and the leaves of many plants,
designed to aid run-off of rain water. Similar surface patterning have
also been exploited in the design of microfluidic devices. For example,
Zhao {\it et\ al.\ } \cite{Zhao1,Zhao2} have used chemically patterned stripes inside
microchannels to control the behaviour of liquid streams and Handique
{\it et\ al.\ } \cite{Handique1} have used a hydrophobic surface treatment
to construct a device which can deliver a fixed volume of fluid.
Our aim here is to use numerical simulations to further demonstrate ways
in which chemically patterned surfaces can be used to manipulate the behaviour of
micron-scale drops, demonstrating that they provide a useful tool
for exploring device designs.

One application where one might envisage a use for surface patterning
is in controlling the size and polydispersity of an array of
drops, an important consideration in many microfluidic
technologies, e.g. \cite{Link1,Squires1,Kuksenok1,Kwok1,Dittrich,Fu,Wang,Ahn1,Cristini1,Pollack,Schwartz,Sugiura,Calvo,Anna}
and the references therein. We explore numerically two ways in which this might be possible. The first uses 
hydrophilic stripes of different widths to sort drops by size or wetting properties. 
The second produces monodisperse drops by exploiting their tendency to remain pinned 
on hydrophilic regions of the surface.

Most drop sorting techniques available in the literature involve active actuation to
control the motion of the drops. This can be done by (but is not limited to) 
electro-osmotic \cite{Dittrich}, mechanical \cite{Fu}, optical \cite{Wang}, or 
dielectrophoretic \cite{Ahn1} manipulation. Our proposed devices are passive 
devices, which require neither a detection nor switching mechanism. Tan {\it et\ al.}
\cite{Cristini1} make use of the channel geometry to control the flow and hence
sort drops by their size. Here we propose a different drop
sorting device which uses chemical surface patterning to separate the liquid drops. 
Since a given surface is wet differently by different liquids, our proposed
method can be used to separate drops based on their wetting properties, size, or both.

Active \cite{Pollack,Schwartz} and passive \cite{Sugiura,Calvo,Anna} drop generating 
techniques are known in the literature, but to the best of our knowledge, they are
mostly concerned with generating liquid drops from continuous streams or larger 
sample drops. Here, we look at the problem from a different angle and 
suggest how monodisperse liquid drops can be formed from small polydisperse drops. 

Since capillary phenomena are increasingly important for systems with large 
surface to volume ratio, passive microfluidic devices based on surface chemical
patterning can play an important role in the future, particularly as the devices 
are downscaled. Careful designs of capillary networks may even lead to an
integrated device, where drop generation, sorting, mixing, and delivery can all
be done on a single chip. We further note that it is an intriguing idea to
incorporate electrowetting into both devices, adding active control, flexibilities,
and functionalities to the devices. Such devices are not limited to the liquid--
gas system considered in this paper. The designs can be used for both a liquid--
liquid and liquid--gas system, and in both an open and closed channel.

We use a lattice Boltzmann algorithm \cite{Briant1,Swift1,Succi1} to solve the 
hydrodynamic equations of motion for the drop movement across the patterned 
substrates thus exploring the key parameters of possible devices. This 
numerical approach has been shown to agree well with experiments in several 
cases \cite{Kusumaatmaja1,Leopoldes1,Dupuis1}, giving us confidence in using
it for the more complicated situations considered here.

\section{Equations of Motions}
The equilibrium properties of the drop are described by a continuum free energy
\begin{equation} 
\Psi = \int_V (\psi_b(n)+\frac{\kappa}{2} (\partial_{\alpha}n)^2) dV
+ \int_S \psi_s(n_s) dS . \label{eq3}
\end{equation}
$\psi_b(n)$ is a bulk free energy term which we take to be \cite{Briant1}
\begin{equation}
\psi_b (n) = p_c (\nu_n+1)^2 (\nu_n^2-2\nu_n+3-2\beta\tau_w) \, ,
\end{equation}
where $\nu_n = {(n-n_c)}/{n_c}$, $\tau_w = {(T_c-T)}/{T_c}$ and $n$, $n_c$, $T$, $T_c$ and $p_c$ 
are the local density, critical density, local temperature, critical temperature and critical 
pressure of the fluid respectively. $\beta$ is a constant typically chosen to be 0.1. This 
choice of free energy leads to two coexisting bulk phases of density $n_c(1\pm\sqrt{\beta\tau_w})$. 
The second term in Eq.\ (\ref{eq3}) models the free energy associated with any interfaces in the 
system. $\kappa$ is related to the liquid--gas surface tension via 
$\sigma_{lg} = {(4\sqrt{2\kappa p_c} (\beta\tau_w)^{3/2} n_c)}/3$ \cite{Briant1}. 
The last term describes the interactions between
the fluid and the solid surface. Following Cahn \cite{Cahn} the
surface energy density is taken to be $\psi_s (n) = -\phi \, n_s$,
where $n_s$ is the value of the fluid density at the surface, so that
the strength of interaction, and hence the contact angle, is parameterised by the variable $\phi$. 

The dynamics of the drop is described by the continuity and the Navier-Stokes equations
\begin{eqnarray}
&\partial_{t}n+\partial_{\alpha}(nu_{\alpha})=0 \, , \label{eq1}\\
&\partial_{t}(nu_{\alpha})+\partial_{\beta}(nu_{\alpha}u_{\beta}) = 
- \partial_{\beta}P_{\alpha\beta} + na_{\alpha}  + \nu \partial_{\beta}[n(\partial_{\beta}u_{\alpha} + \partial_{\alpha}u_{\beta} + 
\delta_{\alpha\beta} \partial_{\gamma} u_{\gamma}) ] \, , \label{eq2}
\end{eqnarray}
where $\mathbf{u}$, $\mathbf{P}$, $\nu$, and $\mathbf{a}$ are the local velocity, pressure tensor, 
kinematic viscosity, and acceleration respectively. We impose no-slip boundary conditions 
$\mathbf{u}=0$ on the surfaces.

The thermodynamic properties of the drop are input via the pressure tensor $\mathbf{P}$ which 
can be calculated from the free energy
\begin{eqnarray} 
&P_{\alpha\beta} = (\partial_{\beta}n \frac{\partial}{\partial(\partial_{\alpha}n)} - \delta_{\alpha\beta}) (\psi_b(n) - \mu_b n + \frac{\kappa}{2} (\partial_{\alpha}n)^2) \nonumber \\
&P_{\alpha\beta} = (p_{\mathrm{b}}-\frac{\kappa}{2} (\partial_{\alpha}n)^2 - \kappa n \partial_{\gamma\gamma}n)\delta_{\alpha\beta} + \kappa (\partial_{\alpha}n)(\partial_{\beta}n) \, ,
\end{eqnarray}
where $p_{\mathrm{b}} = p_c (\nu_n+1)^2 (3\nu_n^2-2\nu_n+1-2\beta\tau_w)$ and $\mu_b=\partial_n \psi_b|_{n=n_b}=\frac{4 p_c}{n_c}(1-\beta\tau_w)$.

\section{Lattice Boltzmann}
The lattice Boltzmann algorithm is defined in terms of the dynamics of a set of real numbers
which move on a lattice in a discrete time. A set of distribution functions $\{f_i(\mathbf{r},t)\}$, 
is defined on each lattice site $\mathbf{r}$. Each of these distribution functions can be interpreted 
as the density of the fluid at time $t$ that will move in direction $i$. The directions $\{i\}$ are discrete 
and for a three dimensional system, one needs to take at least fifteen velocity vectors $\mathbf{v_\mathit{i}} = c(0,0,0)$, 
$c(\pm 1,0,0)$, $c(0,\pm 1,0)$, $c(0,0,\pm 1)$, $c(\pm 1,\pm 1,\pm 1)$. $c=\Delta{}x/\Delta{t}$ is the 
lattice speed, and $\Delta x$ and $\Delta t$ represent the discretisation in space and time respectively. 
The distribution functions are related to the physical variables, the fluid density $n$ and its momentum 
$n\mathbf{u}$, through
\begin{equation}
\sum_i f_i = n,   \;\;\;\;\;\;
\sum_i f_i v_{i\alpha} = nu_{\alpha}. 
\end{equation}
Taking a single-time relaxation approximation, the evolution equation for a given distribution 
function $f_i$ takes the form 
\begin{equation}
f_i(\mathbf{r}+\mathbf{v_i}\Delta t, t+\Delta t) = f_i(\mathbf{r}, t) + \frac{f_i^{eq}(\mathbf{r}, t)-f_i(\mathbf{r}, t)}{\tau}+nw_{\sigma} v_{i\alpha} g_{\alpha} \, . \label{eq7}
\end{equation}
where $w_{\sigma}=w_1 = 1/3$ if $|v|=c$ and $w_{\sigma}=w_2 = 1/24$ if $|v|=c\sqrt{3}$. 
The relaxation time $\tau$ tunes the kinematic viscosity via $\nu = ((\Delta x)^2(\tau-1/2))/(3\Delta t)$ 
\cite{Briant1}. The last term in Eq. (\ref{eq7}) is the forcing term, which is related to the acceleration 
through $a_{\alpha} = \frac{c^2g_{\alpha}}{\Delta{t}}$. Since $\Delta x$ and $\Delta t$ are typically
taken to be $1$ in simulation units, the notations $a_{\alpha}$ and $g_{\alpha}$ are interchangeable
in this paper.

It can be shown that Eq. (\ref{eq7}) reproduces Eqs. (\ref{eq1}) and 
(\ref{eq2}) in the continuum limit if the correct thermodynamic and hydrodynamic information is 
input to the simulation by a suitable choice of local equilibrium functions, i.e. if the following 
constraints are satisfied
\begin{eqnarray}
&&\sum_i f^{eq}_i = n \, , \;\;\;\;\;\; \sum_i f^{eq}_i v_{i\alpha} = nu_{\alpha} \, , \label{eq9} \\
&&\sum_i f^{eq}_i v_{i\alpha}v_{i\beta} = P_{\alpha\beta} + nu_{\alpha}u_{\beta} + \nu [ u_\alpha \partial_\beta n+u_\beta  \partial_\alpha n + \delta_{\alpha \beta}u_\gamma \partial_\gamma n]\, .
\nonumber \\
&&\sum_i f^{eq}_i v_{i\alpha}v_{i\beta}v_{i\gamma} = \frac{nc^2}{3} [u_{\alpha}\delta_{\beta\gamma} + u_{\beta}\delta_{\gamma\alpha} + u_{\gamma}\delta_{\alpha\beta} ]\, . \nonumber
\end{eqnarray}
A possible choice for $f^{eq}_i$ is a power series expansion in the velocity \cite{Leopoldes1,Dupuis2}
\begin{eqnarray}
&&f^{eq}_i = A_{\sigma} + B_{\sigma} u_{\alpha}v_{i\alpha} + C_{\sigma}\mathbf{u}^2 + D_{\sigma}u_{\alpha}u_{\beta}v_{i\alpha}v_{i\beta} + G_{\sigma\alpha\beta}v_{i\alpha}v_{i\beta} \, , \\
&&A_{\sigma} = \frac{w_{\sigma}}{c^2}(p_b(n)-\frac{\kappa}{2} (\partial_{\gamma}n)^2 -\kappa n \partial_{\gamma\gamma}n + \nu u_{\gamma}\partial_{\gamma}n) \, , \nonumber \\
&&B_{\sigma} = \frac{w_{\sigma}n}{c^2} \, , \;\;\;\;\;\; C_{\sigma} = -\frac{w_{\sigma}n}{2c^2} \, , \;\;\;\;\;\; D_{\sigma} = \frac{3w_{\sigma}n}{2c^4} \, , \nonumber \\
&&G_{1\alpha\alpha} = \frac{1}{2c^4} (\kappa (\partial_{\alpha}n)^2 + 2 \nu u_{\alpha}\partial_{\alpha}n) \, , \;\;\;\;\;\; G_{2\alpha\alpha} = 0 \, , \nonumber \\
&&G_{2\alpha\beta} = \frac{1}{16c^4} (\kappa (\partial_{\alpha}n)(\partial_{\beta}n) + \nu u_{\alpha}\partial_{\beta}n + \nu u_{\beta}\partial_{\alpha}n) \, . \nonumber 
\end{eqnarray}
Full details of the lattice Boltzmann algorithm are given in \cite{Dupuis2,Briant1,Swift1,Succi1}.


\section{Sorting drops by size and wetting properties}
To use chemical patterning to sort drops according to size or
wettability we consider the design in Fig. 1. 
A surface is patterned with a rectangular grid of hydrophilic 
(relative to the background) stripes. A drop is input to the 
device at {\it{A}} and subject to a body force at an angle 
$< 45^o$ to the $x$-axis. The system is confined in a channel 
of height $L_z$.

The path taken by the drop through the device depends on the 
drop contact angles with the substrate and the strength of the 
body force. It also, of particular relevance to us here, depends 
on the width of the stripes relative to the drop radius.  By 
choosing the stripes along the $y$ direction to be of equal widths, but those 
along $x$ to increase in width with increasing $y$, drops of different sizes move 
along different paths with the  larger drops moving further along 
the $y$ direction.

Fig. 2(a) -- (c) show simulations of the paths of
drops of initial radius $R=$ 25, 26 and 29 moving through such a
device. The simulation parameters are:
$\theta_{\mathrm{philic}} = 60^{\mathrm{o}}$,
$\theta_{\mathrm{phobic}} = 110^{\mathrm{o}}$, $L_z = 80$, $a_x
=3.0\,10^{-7}$, $a_y = 2.0\,10^{-7}$, dynamic viscosity $\eta=0.4128$,
$\sigma_{lg}=7.7\,10^{-4}$, drop density $n_l=4.128$, $\delta_1 = 20$, $\delta_2
= 30$, $\delta_3 = 40$, $\delta_V=20$, $\delta_H=80$, and $\delta_S=200$. 
The drop velocity is measured to be $\sim 5 \, . \, 10^{-3}$. Comparing 
the simulation results with experiments for a simpler surface pattern
\cite{Kusumaatmaja1}, we established a mapping between simulation and 
physical units, except that the drops move too fast by a factor
of $\sim 500$ in the simulations. (This is a consequence of the
interface width being too wide relative to the drop size in this,
and indeed all, mesoscale simulations.)  
Taking into account this empirical factor, the parameters used here 
correspond to dimensionless numbers: $Re = n_l v R/\eta \sim 3 \, 10^{-3}$, 
$Ca = \eta v/\sigma_{lg} \sim 6 \, 10^{-3}$, 
$Bo = n_l a R^2/\sigma_{lg} \sim 1.2$, 
and $\beta^2 = n_l v^2 R/\sigma_{lg} \sim 2 \, 10^{-5}$.

When a submilimetric drop of initial radius $R$ is jetted on a flat
homogeneous substrate, it will relax to form a spherical cap with a
contact angle given by the Young's law. This is however not generally
the case on chemically patterned surfaces. The equilibrium drop shape
at {\it{A}} is elongated in the $x$-direction because the
drop prefers to wet the hydrophilic stripe. The introduction of the
body force will further deform the drop shape so that it is no longer
symmetric in either the $x$ or the $y$-directions. Indeed, the drop will be 
confined in a hydrophilic stripe
only if the effective capillary force is able to counterbalance the
imposed body force.

In cases where the drops are confined in the $\delta_1$ stripe, they
will move in the $x$-direction from {\it{A}} to the cross-junction
{\it{B}}, where their paths may diverge. In order for a drop to move
in the $y$-direction, the capillary force in this
direction must be large enough to overcome the sum of the capillary force
and the excess external body force in the $x$-direction (recall $a_x >
a_y$). This is where the asymmetry of the drop shape comes into
play. As the volume of the drop is increased, a larger fraction of it
overhangs the stripes and hence a larger fraction 
will interact with the hydrophilic stripe along the
$y$-direction at the junction. This increases the capillary force
along $y$ and means that larger drops (e.g. $R=$ 26) will move in
the $y$-direction to point {\it{C}},
whereas smaller drops (e.g. $R=$ 25) will continue to move along $x$.
 
Since $a_x > a_y$ and $\delta_2 > \delta_V$ the asymmetry of the
drops' shape guarantees that drops at
{\it{C}} move to {\it{D}}. If the widths of the stripes along $x$ were
all the same the drops would then move from {\it{D}} to {\it{E}} for the
same reason that they moved from {\it{B}} to {\it{C}}. However, increasing
the stripe width so that $\delta_2 > \delta_1$ reduces the effective
capillary force in the $y$-direction and therefore some of the drops 
(e.g. $R=$ 26) continue to move along $x$ whereas larger 
drops are routed around the corner towards the third stripe. 
Hence the drops are sorted by size.

Fig. 2(c) shows the path of a drop of radius $R=$ 29 through the
device. Although it does move up to the third stripe it has to pass
several junctions before it does so. This occurs because the drop takes
time to relax to its steady state shape as it moves along the
$\delta_2$ stripe -- note that the drop morphology is slightly different
as it crosses the second and third vertical stripes, with a slightly
increased overhang at the latter. Ideally the distance between two
vertical hydrophilic stripes should have been longer to overcome this
effect.
 
These simulations suggest that by increasing the number of stripes and carefully controlling their
widths it may be possible to sort polydisperse drops
into collections of monodisperse drops.  The precision with which this
can be done is
determined by the increase in the stripe widths. 
Here the vertical stripe width is fixed. As a result
$(R_{n+1}-R_{n}) > (R_{n}-R_{n-1})$. $ (R_{n+1}-R_{n}) \sim (R_{n}-R_{n-1})$ can be achieved
by increasing the vertical stripe width with increasing $x$. 
Two other parameters, the wettability contrast and the external
body force, could also be adjusted to fine-tune the device.

This design can also be applied to sort drops of similar size that
possess different wetting properties. Returning to
Fig. 1 all the drops will move to the
right from point {\it{A}} to point {\it{B}}. Drops with a lower
contact angle on the stripes will overhang the stripes less. Hence at {\it{B}}, these drops
will interact less with the stripes along $y$ and will continue
to move along $x$. As the contact angle of the drops increases they
will feel more effect from the junction and, at some threshold
wettability, they will turn the
corner to move along $y$. As shown in Fig. 3, we find that for drops of radius $R=26$ the
drop moves along stripe $\delta_1$ for  
$\theta_{\mathrm{philic}} = 40^{\mathrm{o}}$
but  is diverted to stripe $\delta_2$ for $\theta_{\mathrm{philic}} =
60^{\mathrm{o}}$, with other simulation parameters as before. 

The current device speed and throughput are limited by the need to eliminate
unwanted drop coalescence. For example, in the drop sorter, drops of different sizes
and wetting properties move at a different speed. If two drops are
entered too closely together, it may lead to a situation where the two drops coalesce 
and mix instead of being sorted.  


\section{Generating monodisperse drops}
As a second example we demonstrate how suitable chemical patterning can be used
to generate monodisperse drops. In
Fig. 4(a) regions {\it{A}} and {\it{C}} are hydrophobic to the drop
relative to regions {\it{B}} and {\it{D}}. For simplicity, we
take $\theta_A = \theta_C = 110^{\mathrm{o}}$ and $\theta_B =
\theta_D = 60^{\mathrm{o}}$. Polydisperse drops are input to the
system in region {\it{A}} and subject to an external body force in the
$x$-direction. If the applied body force is large, all drops
will exit the system at region {\it{D}} without change. 

If the body force is small, however, the effective capillary force
at the {\it{B}}--{\it{C}} border can balance the external force and
the drops are trapped in the hydrophilic stripe. 
As more and more polydisperse drops coalesce at the {\it{B}}--{\it{C}}
border, the resultant drop will no longer be confined to the
hydrophilic stripe. A fraction of the drop volume will start to occupy
the hydrophobic region. At a certain critical volume, the front end of
the drop will reach the {\it{D}} region and the drop will tunnel from
{\it{B}} to {\it{D}}. This is shown in Fig. 4 where we
have presented the lattice Boltzmann simulation results for the
following set of parameters: $\delta_A = 60$, $\delta_B = 70$,
$\delta_C = 30$, $\delta_D = 140$, and $a_x=10^{-7}$. Other parameters
remain the same.

The accuracy with which it is possible to control the final 
drop size is dependent on the ratio between the
average input drop volume to that of the generated output drop: the
smaller the ratio the better the accuracy. There are three key
parameters that can be used to tune the generated drop volume: the
wettability contrast, the stripe widths, and the external body
force. Increasing either the wettability contrast or the stripe widths
will increase the effective capillary force, and hence increase the
final drop volume. Increasing
the body force will have the opposite effect, resulting in a smaller
output drop.
It is also important to tune the input drop frequency. If the
frequency is too high then a number of input drops will coalesce with
the resultant drop as it moves from region {\it{B}} to {\it{D}}. 


\section{Discussion}

To conclude, we have proposed two ways in which chemically
patterned surfaces can be used to control drop size and
polydispersity. The drop sorter takes a collection of
polydiperse drops as input and separates them into different channels
based on their size or wetting properties. 
The drop generator allows small polydisperse drops to coalesce, pinning the 
resultant drop until it reaches a critical volume.

The behaviour of drops within the devices was studied by solving their
equations of motion using a lattice Boltzmann algorithm. Thermodynamic
properties of the drops, such as surface tension and contact angles
were included by using a simple free energy model: in equilibrium
the drops minimise a Landau free energy functional. 

Our results show the potential of chemical patterning as a way to
control the behaviour of micron-scale drops  and demonstrate that simulations
provide a useful tool for exploring possible device designs.  Further
progress needs experimental input to  help to understand more fully
the mapping between physical and  simulation parameters and the
experimental constraints in the realisation  of these devices. We note
that the simulations are computer-intensive, for  example the results
in Fig. 2(c) took about a month on 8 nodes  of dual
2.8 GHz Xeon machines. Synergy with experiment will help  focus computer
resources on the most realistic and useful questions and parameter
ranges.

Acknowledgements:
We thank Alexandre Dupuis and Julien L\'{e}opold\`{e}s for useful discussions. HK acknowledges support from a Clarendon Bursary.



\clearpage

FIGURE CAPTIONS

Fig. 1: Schematic diagram of a drop sorter. The grey stripes on the
surface are hydrophilic with respect to the background. $\delta$
labels the widths of the stripes and $\underline{a}$ 
the imposed acceleration. The arrows show possible
paths of a drop through the device.

Fig. 2: Paths taken by drops of radius (a) $R=$ 25, (b) $R=$ 26, and (c) $R=$ 
29 through the drop sorter. $\delta_1 = 20$, $\delta_2 = 30$, $\delta_3 = 40$, 
and $\delta_V=20$.

Fig. 3: Paths taken by drops of radius $R=$ 26 through the drop
sorter, when the equilibrium contact angle of the drop on the hydrophilic stripes
is (a) $\theta_{\mathrm{philic}} = 40^{\mathrm{o}}$ and (b) 
$\theta_{\mathrm{philic}} = 60^{\mathrm{o}}$.

Fig. 4: Simulations showing the evolution of the drop shape and position
with time in the drop generator. (a) Contour plot as volume is added to the
system. (b)--(e) Snapshots as volume is added. (b) $V = 1.2 \, 10^{5}$.
(c) $V = 2.6 \, 10^{5}$. (d) $V = 3.3 \, 10^{5}$. (e) $V = 4.1 \, 10^{5}$.


\clearpage
\begin{figure}
\centering
\includegraphics[scale=1.25,angle=0]{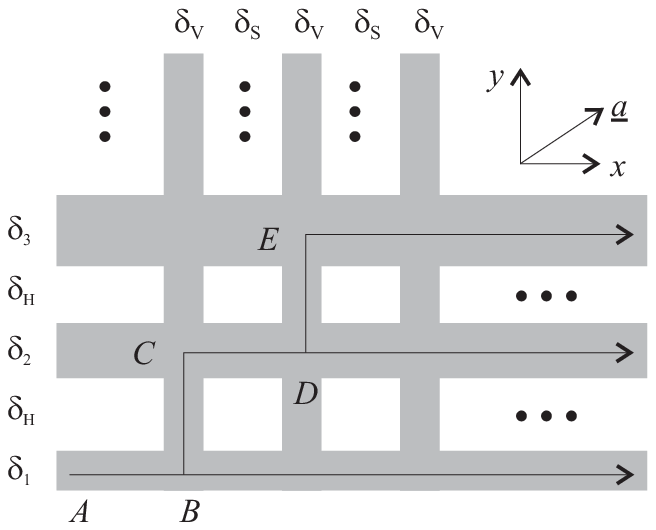}
\caption{
}
\label{fig1:subfig}
\end{figure}

\clearpage
\begin{figure*}
\centering
\includegraphics[scale=1.1,angle=0]{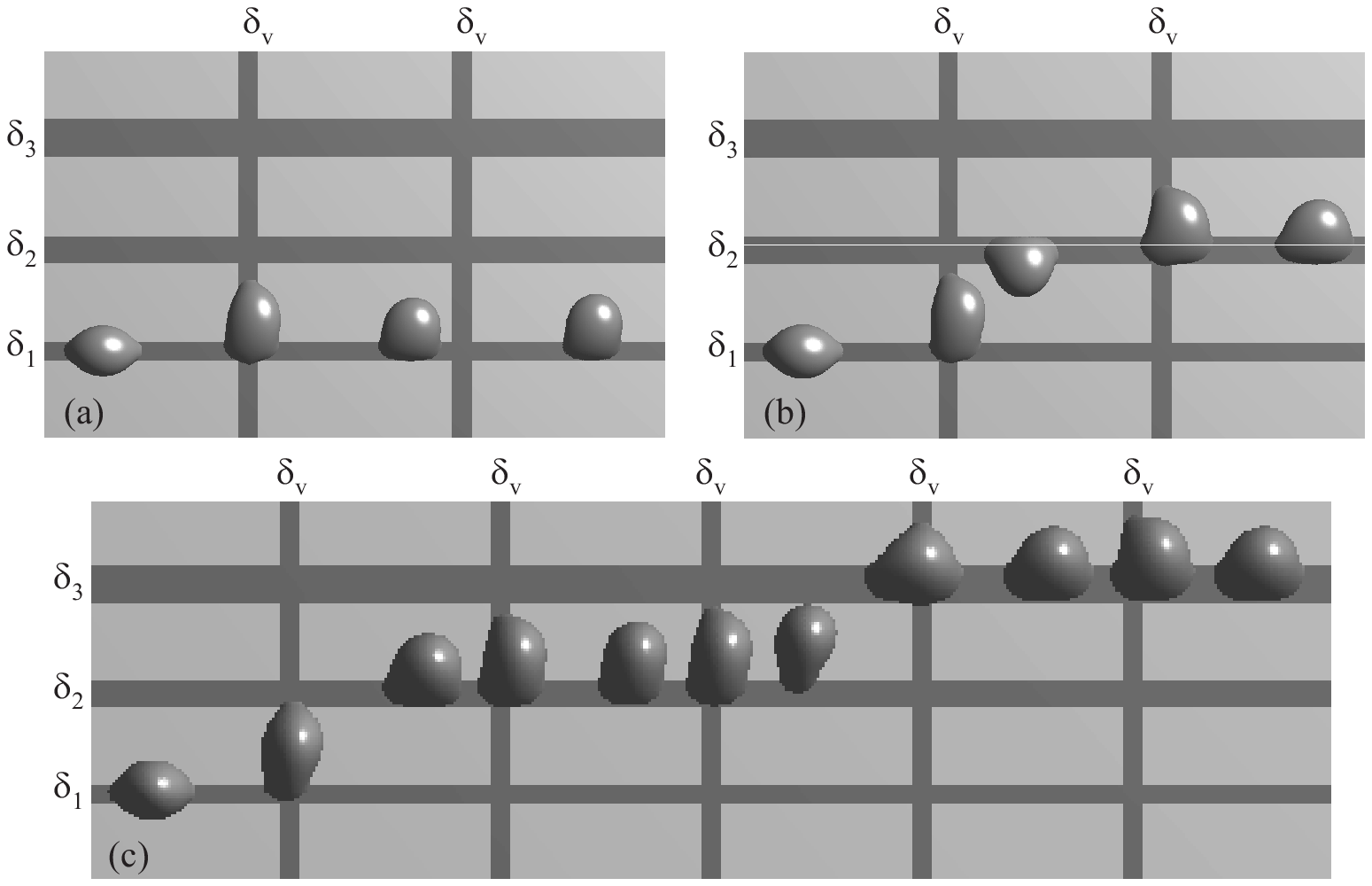}
\caption{
}.
\label{fig2:subfig}
\end{figure*}

\clearpage
\begin{figure}
\centering
\includegraphics[scale=1.1,angle=0]{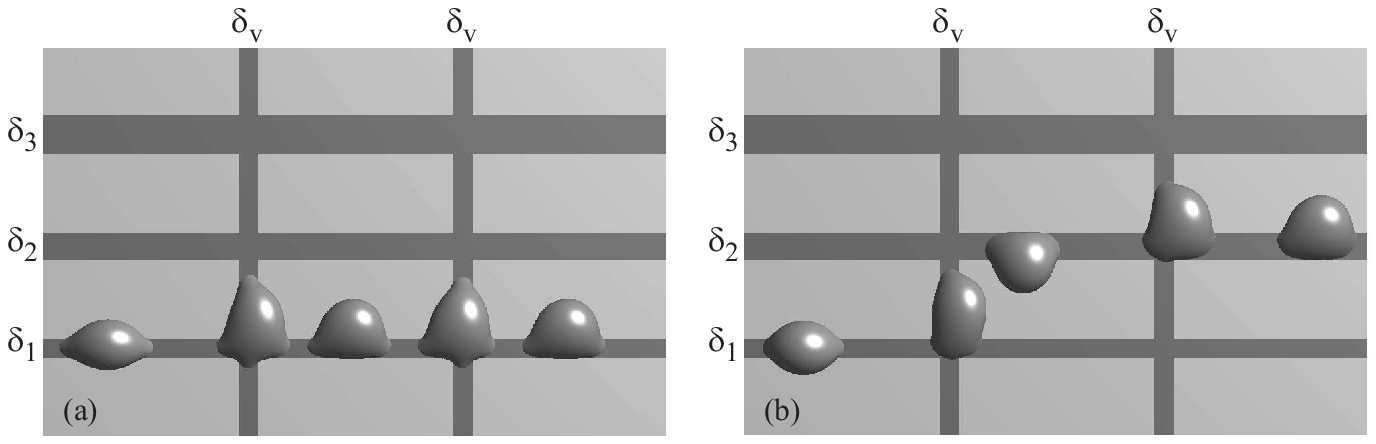} 
\caption{
} 
\label{fig3:subfig}
\end{figure}

\clearpage
\begin{figure}
\centering
\includegraphics[scale=1.25,angle=0]{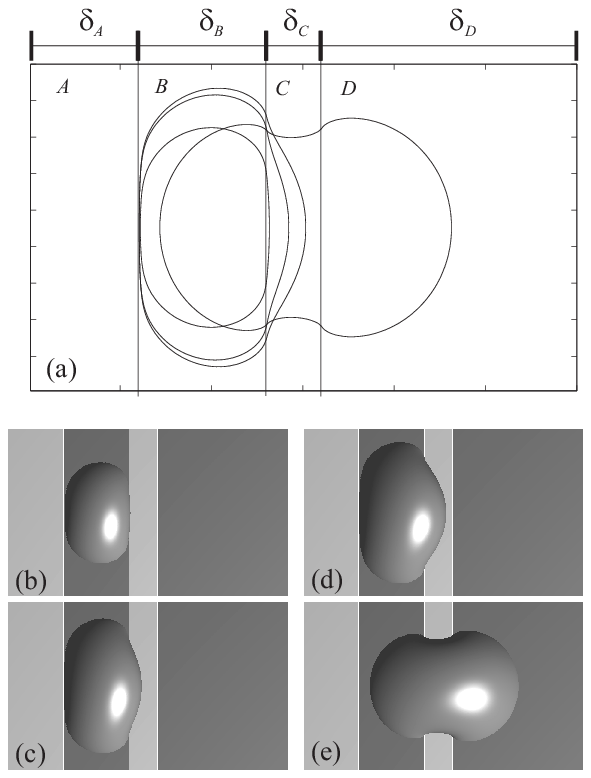}
\caption{
}.
\label{fig4:subfig}
\end{figure}


\end{document}